\documentstyle[epsf,twocolumn,prl,aps,amssymb]{revtex}
\begin{document}
\wideabs
{

\draft
\title{The Polarizability and Electric Field-Induced Energy Gaps In Carbon Nanotubes}
\author{ Xin Zhou$^{1}$, Hu Chen$^{2}$ and Ou-Yang 
Zhong-can$^{1,2}$}
\address{$^{1}$Institute of Theoretical Physics, The Chinese 
Academy of Sciences, P. O. Box 2735, Beijing 100080, China
\\ $^{2}$Centre for Advanced Study, Tsinghua University, Beijing 100084,
 China}
\date{\today}

\maketitle

\begin{abstract}
A simple method to calculate the static electric polarization of single-walled
carbon nanotube (SWNT) is obtained within the second-order perturbation 
approximation. The results are in agreement with the previous calculation 
within the random-phase approximation. We also find that
the low energy electronic band structure of SWNT can be affected by 
an external electric field perpendicular 
to the axis of the tube. The field-induced energy 
gap shows strong dependence on the electric field and the 
size of the tubes, and a metal-insulator transition is predicted for 
all $(n, n)$ tubes. Universal scaling is found for the gap as function of the 
electric field and the radius of SWNTs.
The fact that the external field required to induce a $0.1$ ${\rm eV}$ gap in 
SWNTs can be reached under the currently available experimental 
conditions
indicates a possibility of further applying nanotubes to electric 
signal-controlled nanoscale switching devices.
\end{abstract}

}
%\narrowline

The prospect of nanoscale electronic devices has engaged great 
interest.
Single-walled carbon nanotubes (SWNTs) are significant application 
as nanoscale devices~\cite{Ptody} 
due to their extraordinarily small diameter and versatile 
electronic properties~\cite{Hamada}. It is suggested that individual SWNT 
may act as
 devices such as field-effect transistors (FETs)~\cite{Tans1}, 
single-electron-tunneling transistors~\cite{Tans2,Bockrath}, 
rectifiers~\cite{Yao,Fuhrer}, or p-n junctions~\cite{Leonard}. The
most exciting expectancy lies in the devices fabricated on a single tube
~\cite{[9]}.
In recent years, the interplay between mechanical deformation and 
electrical properties of SWNTs have been extensively 
studied~\cite{[9],Bezryadin,Crespi,Kane}. 
Tombler {\it et al.}~\cite{Tombler,Liu} used an atomic force
microscope tip to manipulate a metallic SWNT, leading to a reversible 
two-order magnitude change of conductance, and Lammert 
{\it et al.}~\cite{Lammert} applied a uniaxial stress to squash 
SWNTs and detect a similar reversible metal-insulator (M-I) transition. 
It is also well known that a magnetic field can also change the conductance of 
carbon nanotubes~\cite{Ajiki,Lu,Roche}. 
A possible electric field-controlled M-I transition 
 are considered to be more exciting because of its easy implementation 
in the actual applications. Yet, a question remains: 
Can electric field change the electronic properties of a tube?
  
In previous studies on electronic transports~\cite{Datta}, the 
potential of a weak longitudinal electric field (bias voltage) in
conductors was treated approximately to a slowly change variable 
 in the range of the primitive unit cell. 
The electric field makes all the electronic energy and the Fermi
level have a gradient along the field direction, but
the energy-band structure is not change. 
The controlled potential, such as the gate voltage without a drop of 
component in the direction perpendicular to the tube axis in the 
case of FET, is only used to shift the Fermi level or changed the 
carrier concentration~\cite{Tans1}. 
In the literature, according to our knowledge, there is no report 
on using a transverse electric 
field to control the longitudinal electronic transport of 
conductors. 
In $(n,n)$ metallic SWNT, the electrons
nearby Fermi energy are nonlocal in the circumference of the tube since their
circumference-Fermi wavevector is zero~\cite{Hamada}, the classic wave-package 
approximation in slow-change voltage may be not suitable 
in the presence of the strong transverse electric field. 
The ${\rm V/ nm}$ order electric field is enough to 
 obviously break the rotational symmetry about the tube 
 axis, and create new interband and intraband coupling, 
 which may change the low energy electronic 
 properties of SWNTs, and hence affect the electronic transport. 
In the other hand, the field is still less than the $2 \sim 3$ orders of the
atomic interior electric field, can be treated as perturbation.

 In this Letter, we first report the result by using a tight-binding (TB) 
model to  calculate the polarizability of SWNTs in the application of an 
external electric field perpendicular to the tube 
  axis.
The calculated polarizability of SWNT
is in agreement with the previous results within the random-phase 
approximation (RPA)~\cite{Benedict}. 
Then we calculate the low energy electronic structure of SWNTs in the 
electric field. 
The results show obviously valuable effects: 
 (1) The electric field can always induce an energy gap in $(n, n)$ 
metallic SWNTs; 
 (2) There is a maximum gap strongly depended on the 
 radius of the tubes;
 (3) Universal scaling is found for the gap as 
 a function of the field and the size of the tubes, and the numerical 
results are testified by the second order perturbation calculations. 
Our results indicate that the magnitude of the electric field required to 
induce a sizable energy gap in metallic SWNTs
falls into the range of currently available experimental conditions.

 In density function theory, single electronic Kohn-Sham Hamiltonian is
\begin{eqnarray}
  {\cal H}=T + V_{KS}[ \rho( {\mathbf r})] + V_e ( {\mathbf r} ) ,
\label{ham1}
\end{eqnarray}
where $V_{KS} [\rho( \mathbf r)]$ is the effective potential, 
self-consistently depended on the electronic density 
$\rho({\mathbf r})$, hence 
the external electric field ${\mathbf E}_e$. $V_e ( {\mathbf r} )$ is 
the electrostatic potential of $\mathbf E_e$.
$V_p ({\mathbf r})=V_{KS}[\rho({\mathbf r})]
-V_{KS}[\rho_{0}({\mathbf r})]$ can be acted as the
contribution of the polarized charge, where $\rho_{0}$ is the unperturbed 
electronic density. So we can rewritten (\ref{ham1}) as
\begin{eqnarray}
{\cal H}= T+V_{KS}[\rho_{0}( {\mathbf r})] + V({\mathbf r})
\label{ham2}
\end{eqnarray}
where $V({\mathbf r})=V_e({\mathbf r}) + V_p({\mathbf r})$ is the total 
perturbed potential. We have
\begin{eqnarray}
V({\mathbf r}) = V_e({\mathbf r}) +\int{\frac{\rho_{b}({\mathbf r}^{\prime})}
{|{\mathbf r}-{\mathbf r}^{\prime}|} d{\mathbf r}^{\prime} }
+\Delta V_{xc}({\mathbf r})
\label{potential}
\end{eqnarray}
where $\rho_{b}({\mathbf r}) = \rho ({\mathbf r}) - \rho_{0}({\mathbf r})$ 
is the 
polarized charge density, $V_{xc}$ is the exchange-correlation potential. If 
$V({\mathbf r})$ or $V_e({\mathbf r})$ is very small, using the second-order 
perturbation theory, we know 
$\Delta U=\frac{1}{2} 
\int{\rho_{b}({\mathbf r}) V({\mathbf r}) d {\mathbf r} }$ and 
\begin{eqnarray}
&&\Delta U=\frac{1}{2} \int{G({\mathbf r},{\mathbf r}^{\prime}) V({\mathbf r}) 
V({\mathbf r}^{\prime}) d {\mathbf r} d {\mathbf r}^{\prime}},
\label{energy2} \\
&&\Delta V_{xc} ({\mathbf r})=\int{ \frac{ \delta V_{xc}[\rho ({\mathbf r})]}
{\delta \rho_{0} ({\mathbf r}^{\prime}) } \rho_{b} ({\mathbf r}^{\prime}) 
d {\mathbf r}^{\prime} },
\label{exchange}
\end{eqnarray}
where $\Delta U$ is the change of total energy of electrons, 
$G({\mathbf r},{\mathbf r}^{\prime})$ can be gained from the unperturbed 
electronic wave functions $\Psi_{i}({\mathbf r})$ and the unperturbed 
energy levels $E_{i}$, 
\begin{eqnarray}
G({\mathbf r},{\mathbf r}^{\prime})=\sum_{i=occ}\sum_{j=unocc} \frac{1}
{E_i-E_j} [\rho_{ij}({\mathbf r}) \rho^{*}_{ij}({\mathbf r}^{\prime})+c.c]
%+ \rho^{*}_{ij} ({\mathbf r}) \rho_{ij}({\mathbf r}^{\prime})
\end{eqnarray}
where $\rho_{ij}({\mathbf r})=\Psi^{*}_{i}({\mathbf r}) \Psi_{j}({\mathbf r})$,
or,
\begin{eqnarray} 
 G({\mathbf r},{\mathbf r}^{\prime})=  
\frac{ \delta^2 \Delta U }
{ \delta V({\mathbf r}) \delta V({\mathbf r}^{\prime})}.
\label{green}
\end{eqnarray}
  
For an approximation, we treat the $V(\mathbf r)$ of 
(\ref{potential}) as a electrostatic potential. If we suppose the 
polarization in SWNT is uniform and due to the induced-surface-charge 
distribution under a constant external field 
${\mathbf E}_e$~\cite{Benedict}, we have $ E =  E_e - \frac{2 p}{R^2}$,
where $p = \alpha_{0} E$ is the dipole moment per length and R being the 
radius of tubes. By selecting a set basis, using slater $X_{\alpha}$ 
local exchange-correlation approximation, we have numerically solved 
(\ref{potential}) in small tubes with the calculated TB-electron-energy 
change $\Delta U$. 
We found the nonhomogenous effect is very small~\cite{zhoux}. 
Here we will only report the results within the uniform polarization 
approximation.  
From $\alpha_0=-\frac{\delta ^2 \Delta U}{\delta E^2}$, we gain the dielectric 
function $\epsilon$,
\begin{eqnarray}
\epsilon = E_e/E = 1 + 2 \frac{\alpha_{0}}{R^2}. 
\label{epsilon}
\end{eqnarray}
Considering the screened electrons are not confined to the 
surface containing the ions in tubes, after calculating the $\alpha_0$ 
within RPA,
Benedict {\it et al.} replaced the $R$ of (\ref{epsilon}) with
$R_{eff}=R+\delta R$ for calculating $\epsilon$. Fitting the result
of $C_{60}$, they gave $\delta R =1.2 \AA$ for some small radius tubes. 
If Directly starting from the exactly unperturbed wave functions 
and energy levels of tubes, we can gain the real distribution of 
polarized charges. 
By numerically solving (\ref{potential}), (\ref{exchange}) and 
(\ref{green}), we found that the exchange-correlation potential maybe 
decrease partly the depolarized field which is completely attributed to  
$R_{eff}$ by Benedict {\it et al}. However, the order of magnitude of 
$\epsilon$
which we need can be identically obtained from one of both 
(\ref{potential}) and (\ref{epsilon}). 

 The nearest-neighbor TB Hamiltonian has been used successfully for 
calculating the electronic structure of graphite sheet 
and nanotubes~\cite{Hamada}. Our previous works~\cite{Zhou1} show the 
model can well describe the electronic perporties and total energy. 
The Hamiltonian is dependent on the matrix element of potential 
$V(\mathbf r)$,
\begin{eqnarray}
 <{\phi}_{i}({\mathbf r}-{\mathbf R}_{l})|{\cal H}
 |{\phi}_{j}({\mathbf r}-{\mathbf R}_{k})> = {\cal H}_{0}^{ij} (l,k)+ 
 V_{ij}(l,k), 
\end{eqnarray}
where $|{\phi}_{i}({\mathbf r}-{\mathbf R}_{l})>$ is the $i$th ($2s$ or
$2p$) orbit of the ${\mathbf R}_{l}$ carbon atom, ${\cal H}_{0}^{ij}(l,k)$ are 
the Slater-Koster TB parameters in the absence of electric field~\cite{Zhou1}.
Expanding the $V({\mathbf r})$ near the lattice vector ${\mathbf R}_{l}$, 
\begin{eqnarray}
V({\mathbf r}+{\mathbf R}_l) \approx V({\mathbf R}_{l})-
{\mathbf E}({\mathbf R}_{l}) 
\cdot {\mathbf r}, 
\end{eqnarray}
where we suppose the $\mathbf E$ is uniform or 
 slowly change in the size of atom.
Only considering the nearest-neighbor contribution, we have
\begin{eqnarray}
V_{ij}(l,l)&\approx&\delta_{ij} V({\mathbf R}_l)-
{\mathbf E} ({\mathbf R}_{l}) \cdot 
{\mathbf P}_{ij}(0), \\
V_{ij}(l,k)&\approx&s_{ij} V({\mathbf R}_{lk}^{c}) - 
{\mathbf E}({\mathbf R}_{lk}^c) \cdot {\mathbf P}_{ij}(\mathbf d),
\end{eqnarray}
where ${\mathbf P}_{ij}(0)$ and ${\mathbf P}_{ij}({\mathbf d})$ are the dipole 
matrix elements, and $s_{ij}$ are overlap integrals which are only slightly 
affect the electronic energy bands, 
${\mathbf d}={\mathbf R}_{l}-{\mathbf R}_{k}$, and 
${\mathbf R}_{lk}^{c}=({\mathbf R}_{l}+
{\mathbf R}_{K})/2$ is the position of mass centre of $l$ atom and $k$ atom.
Benedict {\it et al.}~\cite{Benedict} showed that the ${\mathbf P}_{ij}$
can be neglected except the 
${\mathbf P}_{sp}(0)=R_{sp} {\hat x} \approx 0.5 {\hat x}$ ${\rm \AA}$. 
From the Hamiltonian, we can calculate the total electronic energy in 
presence of any total electric field ${\mathbf E} ({\mathbf r})$. Since the 
polarizability of $\sigma$ electrons is smaller than that of $\pi$, 
we only calculate the contribution of $\pi$.
The results are shown at Fig.\ref{fig1}. The values of $\alpha_0$ of a 
few small
radius tubes are in agreement with the previous results~\cite{Benedict}, but
$\alpha_0 \approx 2.4 R^{2.4}$ is faster increase as the radius increase 
than the result of Benedict {\it et al.} ($\alpha_0 \approx 2.6 R^2$). 
Fig.\ref{fig1}(b) shows
the obtained value of $\epsilon$ of some $(n, n)$ tubes using the $R$ and 
$R_{eff}$, respectively. 
The $\epsilon$ is about order of $5 \sim 15$ for a few small tubes, slowly 
increases to about $20$ for $(60, 60)$ tube. 
Since the polarization is nearly constant in SWNT, we will only calculate the 
effects of a uniform total electric field in this letter. Fig.\ref{fig2} 
shows the electronic energy bands of a $(10, 10)$ tube. 
When $V_0=0.5$ ${\rm V}$, where $V_0=E R$, a sizable gap about 
$0.3$ ${\rm eV}$ is 
 found at $K_0$, the Fermi wavevector in zero field. As $E$
 increases, the gap increases, and when $V_0=1.5$ 
 ${\rm V}$, the bands structure is obviously deformed. 
 It is surprising to find that the gap decreases as  
 $E$ increases further. When $V_0=3.0$ ${\rm V}$, 
 the zero gap is found, but the Fermi point dramatically moves from $K_0$. 
 
 To probe the above effect in general, we performed the computation 
 for a series of $(n,n)$ tubes. Fig.\ref{fig3} shows  
 the gap as a function of the applied field $V_0$ in $(n, n)$ 
 tubes, where $n$ is from $5$ to $15$. From the figure, we find the 
 determined effect: The electric field can always induce a gap 
 in $(n, n)$ tubes, 
 and the size of the gap strongly depends on the amplitude of 
 the transverse field and
 the tube parameter $n$. For any $(n, n)$ tubes, the gap first 
 increases with increasing field, and reaches a maximum 
 value $E_{gm}$ at the $V_{0m}$, 
 then drops again. Both the 
 maximum gap $E_{gm}$ and the corresponding  
 $V_{0m}$ are approximately proportional to 
 $1/n$, and hence inversely proportional to the radius of tubes, i.e.,
 \begin{eqnarray}
  E_{gm} \approx 6.89\ {\rm eV}/n, 
  {\epsilon}_{m} \approx 12.09\ {\rm eV}/n. \label{em}
 \end{eqnarray}
 The finding shown in (\ref{em}) that electric 
 field effects fastly decrease as the 
 size of conductor increases might be the reason why people have 
 not yet recognized the effect in previous studies. 
 The field dependence of the gap is quite similar for tubes with
 various cross-sectional radii, which invokes us to scale both $E_g$ 
 and $V_0$ up $n$ times their original values.
 The obtained results are shown in Fig.\ref{fig4}. From it we do find 
 the scaled gap to be a universal function of the scaled electric 
 field for all $(n,n)$ tubes. In the low field range,
 for all calculated eleven $(n, n)$ tubes there exists a simple relation:
  $n E_g = \lambda (n V_0)^2$,
 where ${\lambda}$ is a constant, about $0.07$ ${\rm (eV)^{-1}}$.
 In the higher field range, except for a few small-radius tubes 
 such as $(5, 5)$ and $(6, 6)$ tubes, the universal scaling law still holds. 
 
 To understand the above scaling relation, we use perturbation
 theory to calculate the field-induced gap in low field limit. 
 The first-order perturbation approximation only causes shift in the 
Fermi level, showing no contribution to the gap change. 
Calculating up 
to the second order perturbation 
 at $K_0$ point, we obtained the following analytic result
 \begin{equation}
  E_g \approx \frac{\sqrt{3}}{2 \pi h} n E^2 R^2,
  \label{perturb}
 \end{equation}
 where $h$ (=$3.033$ eV) is the hopping parameter in the absence of 
 the electric field~\cite{Zhou1,White}. The contribution of the 
 overlap integral $s$, which is very small, is neglected. Obviously, 
 the second-order perturbation calculation gives almost the same 
 scaling relation as the numerical results in the low field, though 
 the obtained $\lambda \approx 0.09$ ({\rm eV})$^{-1}$ is slightly 
 larger than the numerical result $0.07$ ({\rm eV})$^{-1}$. In the 
 high field range, since the Fermi wavevector is moved from $K_0$, 
 the perturbation theory becomes not suitable.
 However, the low field range may be more compatible with the 
 practical application. In order to open a $0.1$ ${\rm eV}$ gap in 
 the energy bands of $(n, n)$ 
 tubes, $n$ must be smaller than $n_0=6.89/E_g \sim 68$, and 
 the required electric field is,
 \begin{equation}
 |{\vec E}|=\frac{2 \pi}{3 r_0 e} \sqrt{\frac{E_g}{\lambda}} 
 n^{-\frac{3}{2}},
 \end{equation}
 where $r_0$ (=$1.42$ ${\rm \AA}$) is the bond length of carbon atoms 
in the SWNT. Therefore, for example for a $(10,10)$ tube, the required 
field is about  $5 \times 10^8$ ${\rm V/m}$, and 
 for a $(60,60)$ tube, it is about $3 \times 10^7$ ${\rm V/m}$.
Considering the polarizing, the needed external field is about 
$0.5$ ${\rm V/\AA}$ and $0.06$ ${\rm V/\AA}$, respectively.
Our results are encouraged by an important fact that the perturbation
approximation is suitable only required the small total field ${\mathbf E}$, 
need not the small external field 
${\mathbf E}_e = \epsilon {\mathbf E}$.
 Even though $\epsilon$ is larger, our conclusion is still correct, but needing
stronger external field.
The magnitude of the required external field for inducing a sizable gap in 
tubes with larger radius can be reached by the 
 currently available experimental conditions, we wish the above prediction 
can be checked in near future.
  
In summary, we have proposed an electric field-induced M-I
transition in $(n, n)$ SWNTs for the first time. 
The results support the 
argument that SWNTs can be applied as nanoscale electric 
signal-controlled switching devices.
 
 The authors would like to thank Prof. H.-W. Peng, Prof. Z.-B. Su 
 and Dr. H.-J. Zhou 
 for many discussions on the results. The numerical calculations
 were performed partly at ITP-Net and partly at the State Key Lab. 
 of Scientific and Engineering Computing.

\begin{figure}
\centerline{\epsfxsize=6cm \epsfbox{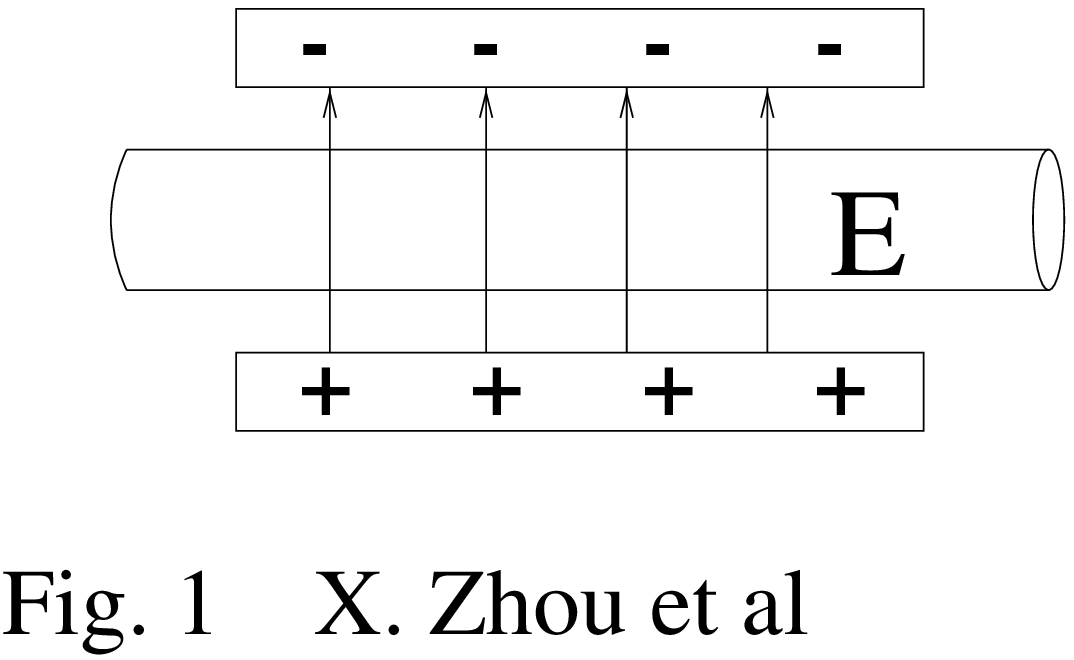}}
\caption{A uniform external electric field perpendicular to the axis of SWNT. 
\label{fig0}}
\end{figure}

\begin{figure}
\centerline{\epsfxsize=9cm \epsfbox{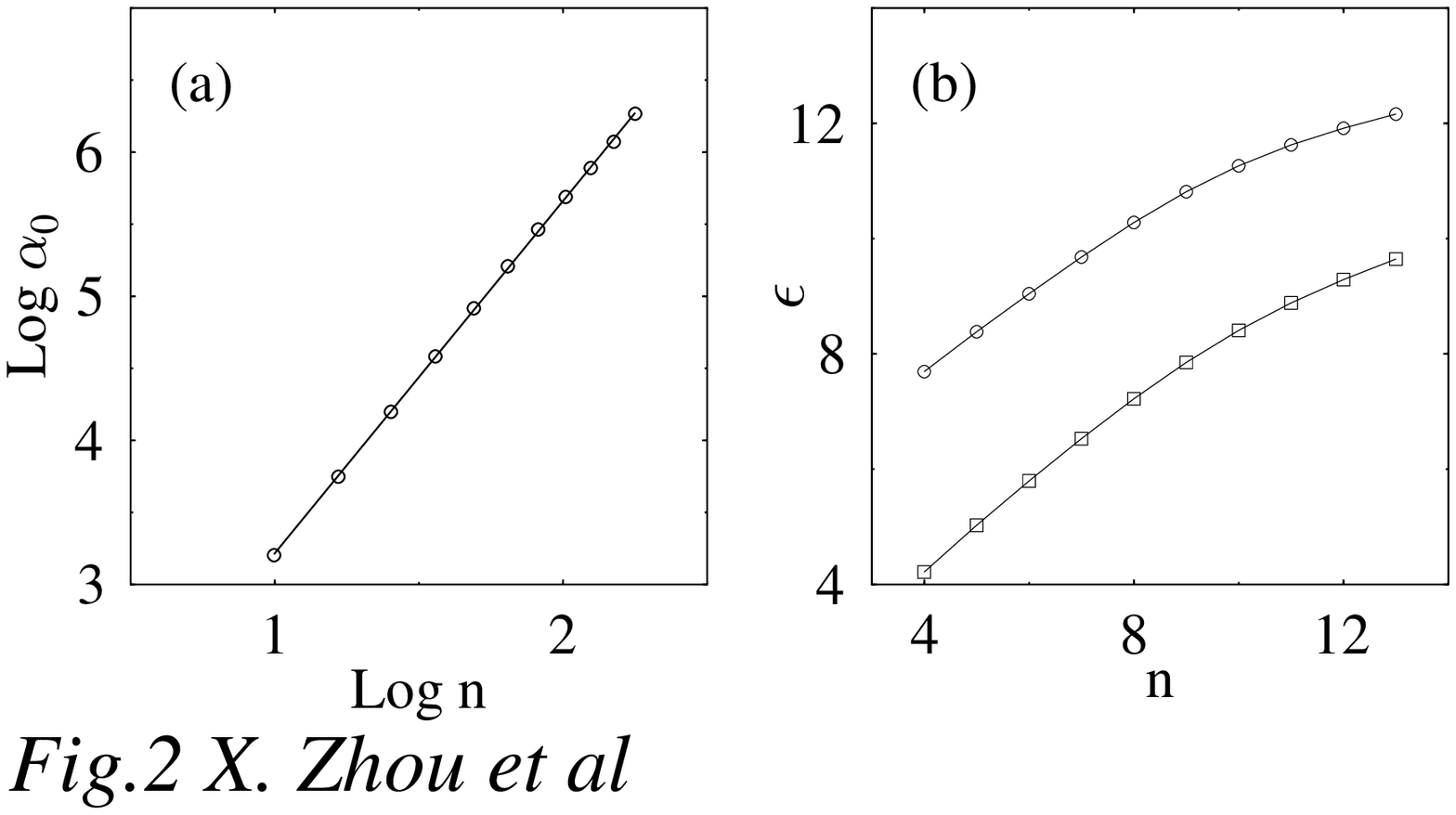}}
\caption{The polarization of $(n, n)$ tubes: (a) $\alpha_0 \sim n^{\gamma}$,
where $\gamma \approx 2.4$ (see text); (b) the dielectric function of tubes, 
circles and squares are the calculated results using $R$ and $R_{eff}$, 
respectively (see text).  
\label{fig1}}
\end{figure}

%\pagebreak

\begin{figure}
\centerline{\epsfxsize=10cm \epsfbox{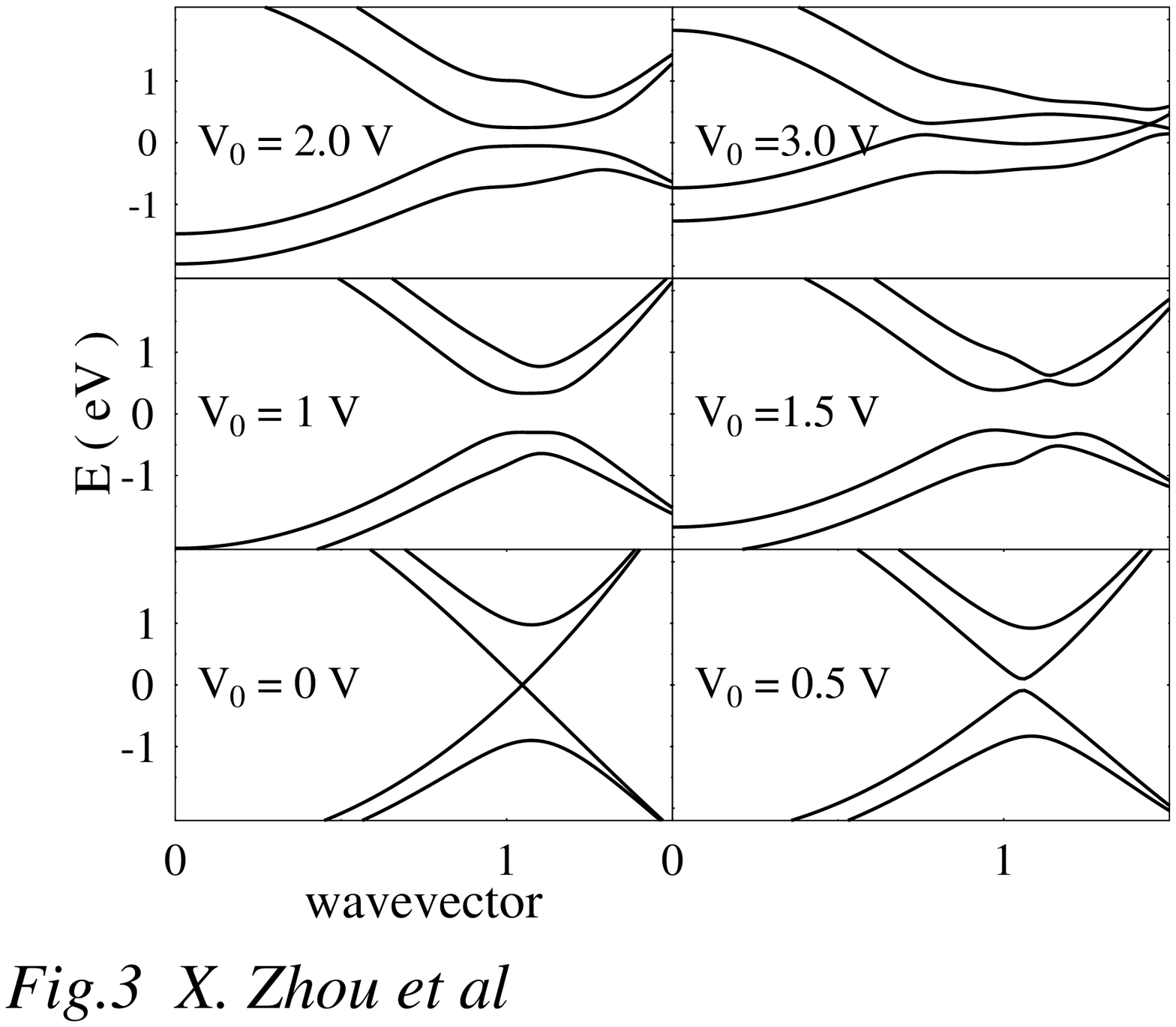}}
\caption{The energy bands of $(10, 10)$ tube in the vicinity of the  
Fermi level under the application of a transverse electric 
field of certain magnitude. $V_0 = 0 V$ is the result in the absence of 
the electric field.
\label{fig2}}
\end{figure}

\begin{figure}
\centerline{\epsfxsize=10cm \epsfbox{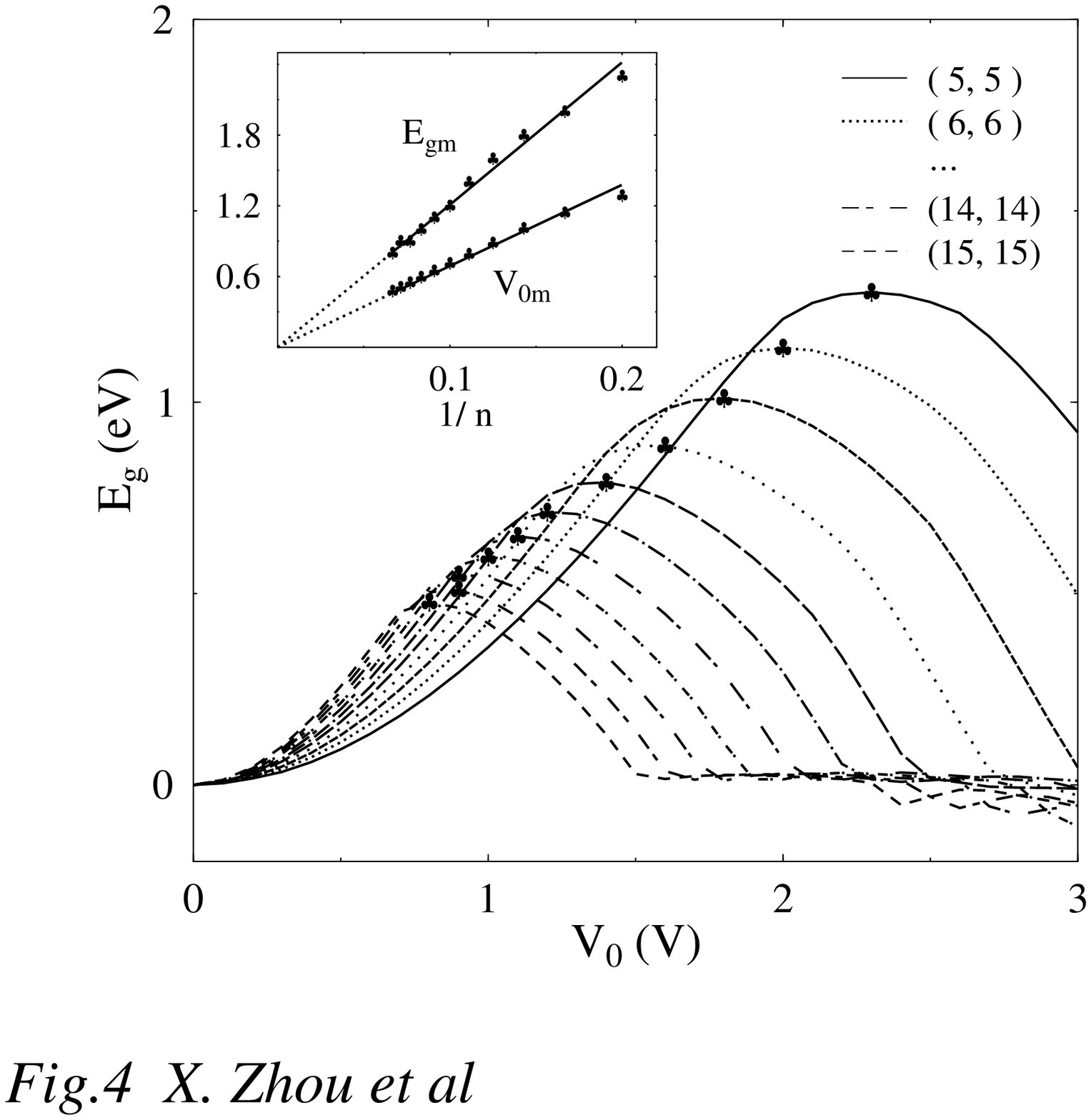}}
\caption{Field-induced gap of $(n, n)$ 
tubes versus the field $V_0 = E R$. From the top to the bottom, 
the tube parameter $n$ increases from $5$ to $15$. The clubs 
denote the position of the maximum gap point. In the inset, both the 
maximum gap $E_{gm}$ and its corresponding electric field $V_{0m}$
are found to be proportional to $1/n$
The lines are fitting results.
\label{fig3}}
\end{figure}

\begin{figure}
\centerline{\epsfxsize=10cm \epsfbox{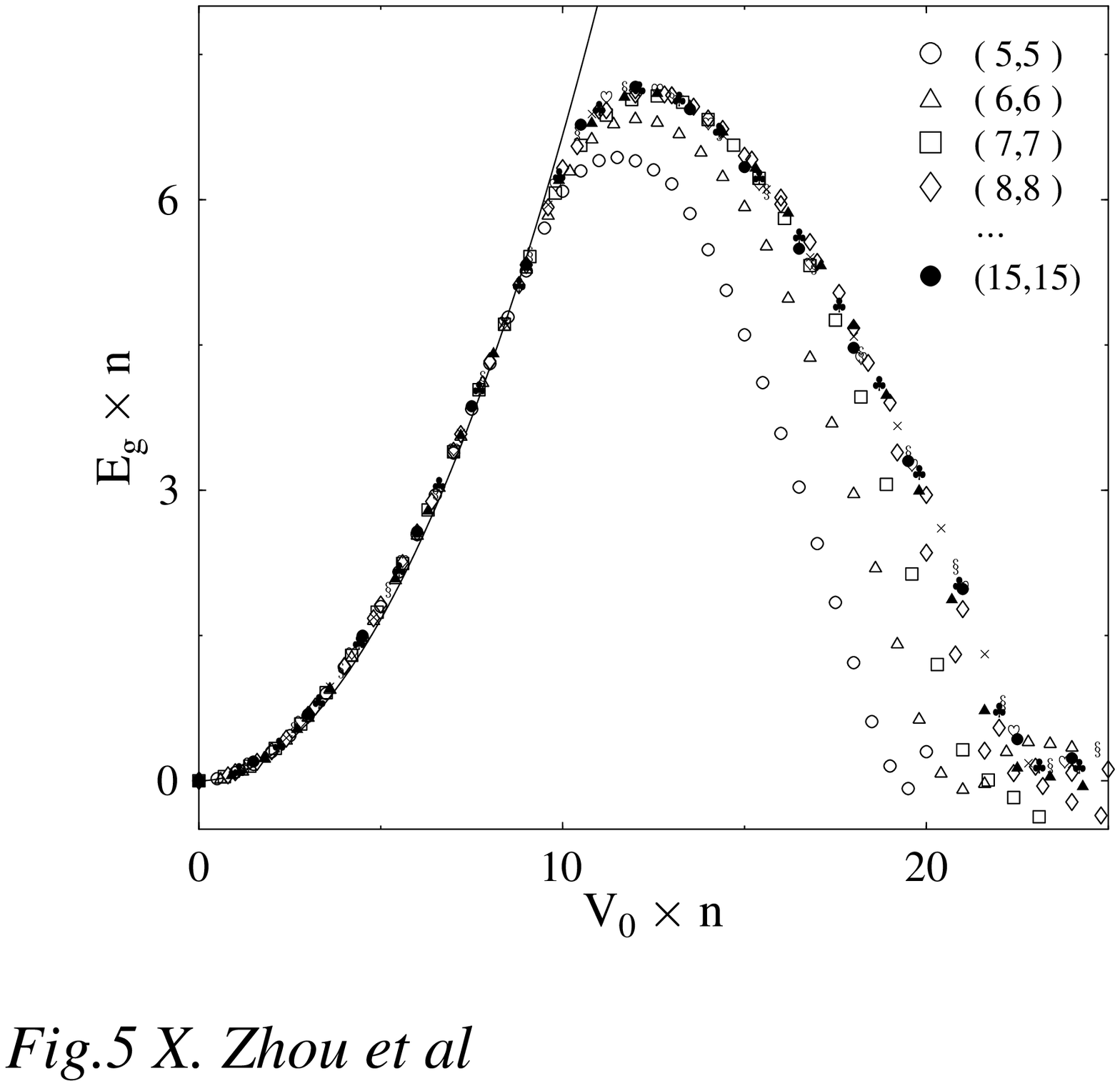}}
\caption{Universal scaling are found in the gap as a
function of $V_0$ in different $(n, n)$ tubes. In the 
low field, the data of all tubes are very much consistent with the scaling 
relation $ n E_g = {\lambda} (n E R)^2 $, as expected by the 
second perturbation theory. The line is the fitting result. 
Except for $(5, 5)$ and $(6, 6)$ 
tubes, the universal scaling is satisfied well up to high field region. 
\label{fig4}}
\end{figure}


\begin{thebibliography}{99}
\bibitem{zhoux} The details will be published in another paper in the future.
\end{thebibliography}
\end{document}